\documentclass[12pt]{article}

\usepackage{amssymb}
\usepackage{amsmath}

\begin{document}

\author{Roman Teisseyre and Mariusz Bia\l ecki \\
\normalsize
\em  Institute of Geophysics, \\ 
\normalsize
\em  
Polish Academy of Sciences, \\ 
\normalsize
\em Ul. Ks. Janusza 64, 01-452 Warszawa, POLAND 
\thanks{ E-mail: {\tt rt@igf.edu.pl}, {\tt bialecki@igf.edu.pl} }}
\title{Complex Relativity: \\
Gravity and Electromagnetic Fields }
\date{}
\maketitle


\begin{abstract}
We present new aspects of the electromagnetic field 
by introducting the natural potentials. These natural potentials are suitable for constructing 
the first order distortions of the metric tensor of Complex 
Relativity - the theory combining the General Relativity with the electromagnetic equations. 
A transition from antisymmetric tensors to the symmetric ones helps to define the natural potentials; 
their form fits a system of the  Dirac matrices and this representation 
leads to distortion of the metric tensor.
    
Our considerations have originated from the recent progresses in the asymmetric continuum theories. 
One version of such theories assumes an existence of the antisymmetric strain and stress fields; 
these fields originate due to some kind of internal friction in a continuum medium 
which have elastic bonds related to rotations of the particles.

\end{abstract}

\bigskip

\section{Introduction}


This work, related to the General Relativity and electromagnetic field, was inspired by some recent 
results in the asymmetric continuum theories including the spin motions; 
therefore, we shall, first of all, quote the numerous attempts to extend the General Relativity 
to include the spin motions. The first one, that in the Cartan works 
\cite{Cartan} was influenced by work by Cosserat brothers \cite{Cosserat} 
in which a moment stress tensor is included in a generalized continuum. 
A gradual development of the Einstein-Cartan Theory (ECT) started by works of  Sciama \cite{Sciama}, 
Kibble \cite{Kibble} and Trautman \cite{Trautman-EC, Trautman-Nature}; 
for a review see: \cite{Hehl}. Kopczyñski \cite{Kopcz73} has proved that in the ECT the cosmological 
solutions become free from the singularities leading to the modified Friedmann equation 
supplemented with the conservation laws for mass and spin \cite{Trautman-ECtheory}.
    
In the XX century, we observed an enormous development of the continuum theories: 
the micropolar and micromorphic theories were developed basing on the Cosserat brothers' work 
(for a review see: \cite{Eringen}); the relations joining the  theories of a continuum
containing defects 
(dislocation and disclination densities) with the Riemannian curvature and torsion were considered by 
Bilby, Bullough and Smith \cite{Bilby55}, Kondo \cite{Kondo55d, Kondo58} and followed by 
Hollander \cite{Hollander62}, Ben-Abraham \cite{Ben-Abr70} and many other authors; 
(for a review see: \cite{Teiss01a, Teiss01b}); 
the thermal stresses were found to have the same form as that related to dislocation field 
\cite{Muskhelishvili} and on this basis the thermal effects were included in the continuum 
with a Riemannian curvature by Kroner \cite{Kroner58, Kroner82}, 
Teisseyre  \cite{Teiss63, Teiss68, Teiss69} and Stojanovic et al. \cite{Stojan64}; 
for a review see: \cite{Teiss01c}).
    
Recently the continuum theories have been generalized to the asymmetic form 
\cite{Teiss02, Teiss05, TeBo03}
in which an additional constitutive law \cite{Shimbo75} for the antisymmetric part of stresses, 
replacing the stress moments, joins the spin motion with a new constant, 
rotation rigidity modulus, to account for the rotation of the point-grains and propagation 
of the spin elastic waves. Such waves can exist when the elastic bonds related to the 
rotation motions of particles are postulated. When , in such a continuum, the material 
bonds for the displacement neglecting originated deformations, there remain only the rotation fields 
of spin and twist types; the respective equations appear to have exactly the form 
of electromagnetic equations. On this basis, the degenerated continuum theory 
(in which there exist only the spin and twist axial motions) has been considered in the last 
papers by Teisseyre \cite{Teiss05} and Teisseyre, Bia{\l}ecki and G{\'o}rski \cite{TeBia05}.
    
Influenced by these results,  we introduce in this paper the natural potentials, 
as defined in the way suitable for constructing the first order distortions in the metric tensor. 
A transition from antisymmetric tensors to the symmetric ones helped us to define these 
natural electromagnetic potentials; their form fits a system of the Dirac matrices and 
this representation leads to distortions of the metric tensor.
    
The definitions helped us to propose a generalization of the General Relativity; 
the new theory - the Complex Relativity - includes, beside gravitation, 
the electromagnetic equations in a first order approximation.

\section{Analogies to Asymmetric Continuum }

We consider the physical rotation fields which can be 
related to the curvature deformations of the complex space. However, we
shall mention that an inspiration for this idea has its source in
considering the rotation and twist motion in the asymmetric elastic
continuum; of course, such motions in an elastic continuum are bounded to
some constitutive relations describing the elastic bonds. This is main
difference in comparison with motion in the space.

In our approach \cite{TeInni03,Teiss04,BoTe04} a homogeneous elastic continuum with
the rotation nuclei - of  spin and twist type - is supplemented, beside the
classical ideal elasticity constitutive law for the symmetric strain-stress 
relation, by the relation between the rotation and asymmetric
stresses; such stresses appear when including in a medium the rotation
nuclei. By this supplementary constitutive law for the anti-symmetric
fields, we can evade an influence of the Hook law, which, when used as the
unique law in the ideal elasticity, rules out an existence of rotation
waves. Thus, it comes out that the rotation vibrations can propagate and are
not attenuated, unlike as the elastic waves in the ideal elastic
continuum.

The twist motion differs from the pure rotation; formaly, it is a
motion composed of the rotation and the mirror reflection. It presents the
simultaneously occurring opposite rotation motions like shear oscillations
(some analogy, in a world of the linear displacements, presents a thermal
expansion/compression motion differing from the simple displacements), but
it can be also related to the axial motions like those of the polarization
type, or, when assuming a posibility of material-space curvature, to
a bending of the 3D space, (by analogy to the situation of a flat
jellyfish with the bending motions (pulsating motions) leading from 2D
form to 3D one).

Our approach to the asymmetry of fields follows from the antisymmetric
stresses introduced by \cite{Shimbo75,Shimbo95} and related to the internal
friction caused by the grain motions under friction forces. 
Note that in the asymmetric continuum also the related asymmetric
incompatibility tensors split into symmetric and anti-symmetric parts.

In our former papers \cite{Teiss01b,Teiss02,TeBo03} we
have also analyzed the theory of asymmetric continuum with defect
distribution (with the dislocation and disclination densities and the
densities of rotation nuclei). Special consideration was paid to rotation
and twist motions related to the definition of the twist-bend tensor.

The dislocation - stress relations and the equations of motion for symmetric
and asymmetric parts of stresses were derived.The obtained relations for
elastic fields, given by difference of the total and self-fields, can be
split into the selfparts prevailing on the fracture plane and the total
parts describing seismic radiation field in a surrounding space. Some
applications were shortly discussed.

Finally, we shall note that a more complex deformation field, like that with
the dislocation and disclination densities, leads, when applying the
material coordinate system, to description of a deformed state in terms of
the Riemannian geometry, or even non-Riemannian one \cite{Bilby55,Kondo55}. 
This remark applies also to thermal deformation field \cite{Teiss69, Teiss01b}.

\section{Natural electromagnetic potentials}

We have tested many variants of definition for the EM potentials, and finally we
propose the following one. We introduce the 3D vector potentials: 
$\tilde{A}_{s}$, $\hat{A}_{s}$, and charge-current potentials $\varphi$, $\psi_s$ 
instead of the standard 4D vector potential $A_{\mu }$. 
We call them the natural potentials and assume they fulfill the following equations:   
\begin{eqnarray} 
B_{k}&=&\epsilon _{kbs}\tilde{A}_{s,b} \ \ \ \ \    \tilde{A}_{s,s}=0  \label{Maxw-Pot1}\\
E_{k}+\varphi_{,k}&=&\epsilon _{kbs}\hat{A}_{s,b} \ \ \ \ \   \hat{A}_{s,s}=0  \label{Maxw-Pot2}\\
\frac{4\pi }{c}J_{k}&=&\frac{1}{c}\dot{\varphi}_{,k}+\frac{1}{c}\epsilon _{kbs}\psi _{s,b}
\label{Maxw-Pot3}\\
\varphi _{,kk}&=&-4\pi \rho, \label{Maxw-Pot4}
\end{eqnarray}
where $k, b, s \in \{ 1,2,3 \}$. An index \emph{after} a comma denotes differentiation and 
the summation convention for repeated indices is used.  
These new natural potentials yield 
\begin{eqnarray} 
\epsilon _{kbs}\tilde{A}_{s,b}-\frac{1}{c}\frac{\partial }{\partial t}
\hat{A}_{k}&=&\frac{1}{c}\psi _{k} \ \ \ \ \    \tilde{A}_{b,b}=0  \label{MaxA1} \\
\epsilon _{kbs}\hat{A}_{s,b}+\frac{1}{c} \frac{\partial }{\partial t}\tilde{A}_{k}&=&0 \ \ \ \ \ 
\hat{A}_{b,b}=0. \label{MaxA2}
\end{eqnarray}
When applying to \eqref{MaxA1}-\eqref{MaxA2} the operator $\epsilon _{ndk}\frac{%
\partial }{\partial x_{d}}$, we arrive at the Maxwell equations:
\begin{eqnarray}
\epsilon _{ndk}B_{k,d}-\frac{1}{c}\frac{\partial }{\partial t}E_{n} &=&\frac{4\pi }{c}J_{k} 
\ \ \ \ \ B_{k,k}=0 \\
\epsilon _{ndk}E_{k,d}+\frac{1}{c}\frac{\partial}{\partial t}B_{n}&=&0 
\ \ \ \ \ E_{k,k}=4\pi \rho 
\end{eqnarray}

Now, we construct the complex antisymmetric tensor $A_{\alpha \beta }$ 
($\alpha, \beta \in \{1,2,3,4\}$) for potentials $\tilde A$, $\hat A$.
We define $A_{\alpha \beta }=\tilde{A}_{\alpha \beta }+\text{i}\hat{A}_{\alpha \beta }$ by 
\begin{equation} \label{tens-def}
A_{\alpha \beta }=\left[ 
\begin{array}{llll}
0 & \tilde{A}_{3} & -\tilde{A}_{2} & -\text{i}\hat{A}_{1} \\ 
-\tilde{A}_{3} & 0 & \tilde{A}_{1} & -\text{i}\hat{A}_{2} \\ 
\tilde{A}_{2} & -\tilde{A}_{1} & 0 & -\text{i}\hat{A}_{3} \\ 
\text{i}\hat{A}_{1} & \text{i}\hat{A}_{2} & \text{i}\hat{A}_{3} & 0%
\end{array}
\right] \ \ \ +\text{i}\left[ 
\begin{array}{llll}
0 & \hat{A}_{3} & -\hat{A}_{2} & \text{i}\tilde{A}_{1} \\ 
-\hat{A}_{3} & 0 & \hat{A}_{1} & \text{i}\tilde{A}_{2} \\ 
\hat{A}_{2} & -\hat{A}_{1} & 0 & \text{i}\tilde{A}_{3} \\ 
-\text{i}\tilde{A}_{1} & -\text{i}\tilde{A}_{2} & -\text{i}\tilde{A}_{3} & 0%
\end{array}
\right] 
\end{equation}
or 
\begin{equation}  \label{defA}
A_{\alpha \beta }=\left[ 
\begin{array}{llll}
0 & \bar{A}_{3} & -\bar{A}_{2} & -\bar{A}_{1} \\ 
-\bar{A}_{3} & 0 & \bar{A}_{1} & -\bar{A}_{2} \\ 
\bar{A}_{2} & -\bar{A}_{1} & 0 & -\bar{A}_{3} \\ 
\bar{A}_{1} & \bar{A}_{2} & \bar{A}_{3} & 0%
\end{array}
\right]  
\end{equation}
where $\bar{A}_{k}=\tilde{A}_{k}+i\hat{A}_{k}$.
The form of equation \eqref{tens-def} follows the fact, that
these potentials are constructed in a similar manner as the tensor $f_{\alpha \beta }$
of the EM field is constructed from the EM vector fields $B_{s}$ and $E_{s}$.

According to \eqref{MaxA1}-\eqref{MaxA2} the tensor $A_{\alpha \beta }$ fulfils the condition 
\begin{equation}
A_{\alpha \beta \left\vert \beta \right. }=\frac{1}{c}\psi _{\alpha }  \, \ \ \ \
\psi _{\alpha }=\{\psi _{k}\text{ , }0\} \, \ \ \ \ k=1,2,3.  
\end{equation}
Comparing this relation with equations for the potentials $\tilde{A}_{s}$, $\hat{A}_{s}$ 
we obtain 
\begin{eqnarray}
\square \tilde{A}_{n}&=&-\frac{1}{c}\epsilon _{ndk}\psi _{k,d} \ \ \ \ \ 
\tilde{A}_{b,b}=0 \label{boxAt} \\
\square \hat{A}_{n}&=&\frac{1}{c^{2}}\frac{\partial }{\partial t}\psi _{n} \ \ \ \ \
\hat{A}_{b,b}=0 \label{boxAh}
\end{eqnarray}
and when defining 
\begin{equation} \label{curpot}
-\frac{1}{c}\epsilon _{ndk}\psi _{k,d}=\tilde{J}_{n}\bigskip \ \ \ \ \ 
\frac{1}{c^{2}}\frac{\partial }{\partial t}\psi _{k}=\hat{J}_{k} 
\end{equation}
we arrive at 
\begin{equation}
\square \tilde{A}_{n}=\tilde{J}_{n}\text{ \ , \ \ \ }\square \hat{A}_{n}=\hat{J}_{n} \label{boxAJJ}
\end{equation}

Then, we define tensor $J_{\alpha \beta }$ : \ 
\begin{equation}
J_{\alpha \beta }=\left[ 
\begin{array}{llll}
0 & \bar{J}_{3} & -\bar{J}_{2} & -\bar{J}_{1} \\ 
-\bar{J}_{3} & 0 & \bar{J}_{1} & -\bar{J}_{2} \\ 
\bar{J}_{2} & -\bar{J}_{1} & 0 & -\bar{J}_{3} \\ 
\bar{J}_{1} & \bar{J}_{2} & \bar{J}_{3} & 0%
\end{array}
\right]
\end{equation}
where $\bar{J}_{n}=\tilde{J}_{n}+$i$\hat{J}_{n}$ and we arrive to relation 
\begin{equation}
\square A_{\alpha \beta }=J_{\alpha \beta }. \label{boxAJ}
\end{equation}

Applying the operator $\epsilon _{sbn}\frac{\partial }{\partial x_{b}}$ we
get, with the help of \eqref{boxAt}-\eqref{boxAh}, 
\begin{eqnarray}
\square B_{s} &=&\frac{1}{c}\psi _{s,dd}\text{ \ }=\frac{4\pi }{c}\epsilon
_{spk}J_{k,p}  \label{MEB} \\
\square E_{s} &=&\frac{1}{c^{2}}\epsilon _{sbn}\frac{\partial }{\partial t}%
\psi _{n,b}\text{\ }-\square \varphi _{,s}=-\frac{4\pi }{c^{2}}\dot{J}%
_{s}+4\pi \rho _{,s}  \ . \label{MEE}
\end{eqnarray}

\section{Natural EM Potential \\--- the Symmetric Tensor}

To include the EM field into the Riemannian or non-Riemannian geometry we
search for a way how to build a metric tensor which in the first
order approximation could describe the gravity and EM fields.

Owing to the fact that for the six potentials $\tilde{A}_{s}$\textbf{\ , \ }$%
\hat{A}_{s}$ we can introduce the two additional conditions, we define the
other set of four potentials: 
\begin{equation}
\tilde{N}_{n}=\{\tilde{N}_{1}\text{ , }\tilde{N}_{2\text{ }},\ 0 \} \ \text{and} \ 
\hat{N}_{n}=\{\hat{N}_{1} , \ \hat{N}_{2}, \ 0 \}. \label{2.1} 
\end{equation}
With these conditions we introduce the natural \textbf{symmetric} tensor of
potentials:
\begin{eqnarray}
N_{\alpha \beta } &=&\tilde{N}_{\alpha \beta }+\text{i}\hat{N}_{\alpha \beta
}\text{\ \ \ \ \  \ \ \ \ \ }N_{s}=\text{\ }\tilde{N}_{s}+\text{i}\hat{N}_{s} \\
N_{\alpha \beta }&=&\left[ 
\begin{array}{cccc}
0 & 0 & -N_{2} & -N_{1} \\ 
0 & 0 & N_{1} & -N_{2} \\ 
-N_{2} & N_{1} & 0 & 0 \\ 
-N_{1} & -N_{2} & 0 & 0 
\end{array}
\right]  \label{2.3} 
\end{eqnarray}

\paragraph{The bridge relations.}

We demand that first bridge relation is satisfied:  
\begin{equation}
\ \epsilon _{kbs}\tilde{N}_{s,b}\ =\epsilon _{kbs}\tilde{A}_{s,b}\ ,\ \
\epsilon _{kbs}\hat{N}_{s,b}=\epsilon _{kbs}\hat{A}_{s,b}\ \label{2.4} 
\end{equation}
and with the conditions 
\begin{equation}
N_{s}\ =\epsilon _{skn}\Omega _{n,k}\text{ , \ }\Omega _{1}=\Theta _{,1}%
\text{ , \ }\Omega _{2}=\Theta _{,2} \label{2.4prime} 
\end{equation}
we have 
\begin{equation}
\text{\ \ }N_{1,1}+N_{2,2}=0\text{ , \ }N_{3}=0. \label{2.4primeprime} 
\end{equation}
The tensor $N_{\alpha \beta} $ (i.e. symmetric EM potential) 
has changed the signs of the terms below the diagonal of the matrix
when comparing  to tensor $A_{\alpha \beta }$ (i.e. antisymmetric EM potential).

To connect the antisymmetric potentials with the symmetric ones we shall
introduce also a new definition for current potential 
\begin{equation}
\eta _{\alpha }=\{\eta _{1},\eta _{2},0,0\} \ , \ \ \ \  \eta _{s,s}=0.  \label{eta:def} 
\end{equation}
We demand the validity of the second bridge relations: 
\begin{equation}
\ \epsilon _{kbS}\eta _{S,b}\ =\epsilon _{kbn}\psi _{n,b}\ \ ; \ \ S=1,2.\
\end{equation}%

\noindent {\bf Remark.} A class of these EM potentials which can be transformed into the form described 
above will be called {\emph{two-component}} EM potentials. We claim such potentials 
have proper form to contribute to metric tensor as described in section \ref{curvature}.

\paragraph{The field relations.}
Having these bridge relations we return to symmetric tensors. Due to this
fact we get  the equations for new
symmetric potentials, similar to \eqref{boxAt}-\eqref{boxAh}, 
\begin{equation}
\square \tilde{N}_{K}=-\frac{1}{c}\epsilon _{KdN}\eta _{N,d}\text{\ \ , \ \ }%
\hat{N}_{S,S}=0\text{ \bigskip ;}\ \text{\ }\square \hat{N}_{K}=\frac{1}{%
c^{2}}\frac{\partial }{\partial t}\eta _{K}\text{\ , \ \ \ }\tilde{N}_{S,S}=0.  \label{2.8prime}
\end{equation}%
Instead of \ relation\ \eqref{curpot} for currents $\tilde{J}_{n}$ \bigskip
and$\ \ \hat{J}_{k}$\ \ we shall introduce a new definition
of currents $\tilde{Y}_{N}$ \bigskip and $\hat{Y}_{N}$ and tensor $Y_{\alpha\beta }:$ 
\begin{equation}
Y_{\alpha \beta }=\left[ 
\begin{array}{cccc}
0 & 0 & -\tilde{Y}_{2} & -\tilde{Y}_{1} \\ 
0 & 0 & \tilde{Y}_{1} & -\tilde{Y}_{2} \\ 
-\tilde{Y}_{2} & \tilde{Y}_{1} & 0 & 0 \\ 
-\tilde{Y}_{1} & -\tilde{Y}_{2} & 0 & 0%
\end{array}%
\right] \ +\text{i}\left[ 
\begin{array}{cccc}
0 & 0 & -\hat{Y}_{2} & -\hat{Y}_{1} \\ 
0 & 0 & \hat{Y}_{1} & -\hat{Y}_{2} \\ 
-\hat{Y}_{2} & \hat{Y}_{1} & 0 & 0 \\ 
-\hat{Y}_{1} & -\hat{Y}_{2} & 0 & 0%
\end{array}%
\right]   \label{2.10}
\end{equation}%
\begin{equation}
-\frac{1}{c}\eta _{1,3}=\tilde{Y}_{2}\text{ \bigskip , \ }\frac{1}{c}\eta
_{2,3}=\tilde{Y}_{1}\text{\ \ ;}\ \ \ \ \ \frac{1}{c^{2}}\frac{\partial }{%
\partial t}\eta _{N}=\hat{Y}_{N}  \label{2.10'}
\end{equation}%
where similarly to \eqref{boxAJJ} we have: 
\begin{equation}
\square \tilde{N}_{N}=\tilde{Y}_{N}\ ,\ \ \ \ \square \hat{N}_{N}=\hat{Y}%
_{N}\   \label{2.11}
\end{equation}%
The following relations for the potentials are to be noted 
\begin{equation}
N_{\alpha \gamma ,\gamma }=\frac{1}{c}\eta _{\alpha }\text{ , \ \ \ }\square
\ N_{\alpha \beta }=Y_{\alpha \beta }  \label{2.13}
\end{equation}

We will show, further on, that all these symmetric matrices are the tensors
related to the Dirac tensors.

\paragraph{Complex metric tensor and EM natural potentials.}
The natural tensor of potentials \eqref{2.3} can be presented as follows:
\begin{equation}
N_{\alpha \beta }=\tilde{N}_{1}\epsilon ^{1}\ +\tilde{N}_{2}\epsilon ^{2}+%
\text{i}\hat{N}_{1}\epsilon ^{1}\ +\text{i}\hat{N}_{2}\epsilon
^{2}=N_{1}\epsilon ^{1}\ +N_{2}\epsilon ^{2} \label{2.15} 
\end{equation}
where 
\begin{equation}
\epsilon ^{1}=\left[ 
\begin{array}{cccc}
 0& 0& 0&-1\\ 
 0& 0& 1& 0\\ 
 0& 1& 0& 0\\ 
-1& 0& 0& 0
\end{array}
\right] ,\ \epsilon ^{2}=\left[ 
\begin{array}{cccc}
0 & 0 & -1 & 0 \\ 
0 & 0 & 0 & -1 \\ 
-1& 0 & 0 & 0 \\ 
0 & -1 & 0 & 0%
\end{array}
\right] 
\end{equation}

The matrices $\epsilon ^{\nu }$ fulfil the conditions for the Dirac's matrices:\ 
\begin{equation}
\epsilon ^{\alpha }\epsilon ^{\beta }+\epsilon ^{\beta }\epsilon ^{\alpha
}=2\eta ^{\alpha \beta }\label{2.17} 
\end{equation}
with  
\begin{equation}
\epsilon ^{3}=\left[ 
\begin{array}{cccc}
0 & 0 & 0 & \text{i} \\ 
0 & 0 & \text{i} & 0 \\ 
0 & -\text{i} & 0 & 0 \\ 
-\text{i} & 0 & 0 & 0%
\end{array}
\right] ,\ \epsilon ^{4}=\left[ 
\begin{array}{cccc}
\text{i} & 0 & 0 & 0 \\ 
0 & \text{i} & 0 & 0 \\ 
0 & 0 & -\text{i} & 0 \\ 
0 & 0 & 0 & -\text{i}%
\end{array}
\right]. \label{2.16} 
\end{equation}
However, when disturbing these matrices in the way indicated below 
\begin{equation}
\gamma ^{1}=(1+N_{1})\epsilon ^{1}\ \text{\ , }\gamma ^{2}=(1+N_{2})\epsilon
^{2}\text{ , \ }\gamma ^{3}=\epsilon ^{3}\text{ \ , \ }\gamma ^{4}=\epsilon^{4}\label{2.18} 
\end{equation}
we obtain the relation for the Dirac's matrices in non-Euclidean space: 
\begin{equation}
\gamma ^{\alpha }\gamma ^{\beta }+\gamma ^{\beta }\gamma ^{\alpha
}=2g^{\alpha \beta }\label{2.19} 
\end{equation}
This relation justifies our approach in which we have assumed
that the natural potentials \eqref{2.3} can be used as the disturbances to metric
tensor in any reference system.

These complex disturbances to metric tensor can be
combined with the gravity disturbances $h_{\alpha \beta }^{G}$ . However
let us first discuss another possible fields constructed similarly
as above.

\bigskip
\noindent{\bf Comment: Field \textbf{V}.}
Thus, our next question is related to what would happen when one 
disturbes the $\epsilon ^{3}$\ and \ $\epsilon ^{4}$ matrices. To this aim,
we will consider the complex tensor potential $V_{\alpha \beta }$ defined as
disturbances to  $\epsilon ^{3},\ \epsilon ^{4}$ similarly as $%
N_{\alpha \beta }$ disturbs $\epsilon ^{1},\ \epsilon ^{2}$ in  \eqref{2.15}: 
\begin{equation}
V_{\alpha \beta }=\text{i}(V_{3}\epsilon ^{3}\ +V_{4}\epsilon ^{4}). \label{2.20} 
\end{equation}
The disturbances related to $V_{3}$ are not symmetric hence we put $V_{3}=0.$

With the undisturbed matrices $\bar{\gamma}^{1}=\epsilon ^{1}$ , $\bar{\gamma%
}^{2}=\epsilon ^{2}$ , $\bar{\gamma}^{3}=\epsilon ^{3}$, we define the
disturbed matrics $\bar{\gamma}^{4}:$%
\begin{equation}
\text{ \ }\bar{\gamma}^{4}=\text{i}(1+V_{4})\epsilon ^{4}\label{2.21} 
\end{equation}
Further on, we obtain the related metric tensor: 
\begin{equation}
\bar{g}_{\mu \nu }=\frac{1}{2}(\bar{\gamma}_{\mu }\bar{\gamma}_{\nu }+\bar{%
\gamma}_{\nu }\bar{\gamma}_{\mu })\label{2.22} 
\end{equation}
Both relations correspond to those given in \eqref{2.18} and \eqref{2.19} 
and, with Einstein equations, can lead us to $ V^{\alpha \beta},_\beta = Z^\alpha$, 
$V^{\alpha \beta},^\gamma_\gamma = U^{\alpha\beta}$ where $Z^\alpha$, $U^{\alpha \beta}$ would
represent some source fields.
 
Further on, we neglect the field $V^{\alpha\beta}$.

\noindent{\bf Remark.} We shall note that, instead of the 4D presentation, it was also
possible to present our relations in the 2D forms with the tensor 
\begin{equation*}
N_{AB}=\left[ 
\begin{array}{ll}
N_{2} & N_{1} \\ 
N_{1} & -N_{2}%
\end{array}%
\right] 
\end{equation*}%
and with the help of the Pauli 2D tensors.

\section{Complex metric and curvature tensors} \label{curvature}

Our formulation of the Complex Relativity is based on the symmetric form of
perturbations introduced into the metric tensor.

The respective metric tensor can be constructed when considering the
coordinates $X^{\alpha }$ ($X^{4}=$i$ct$) for which the numerical values,
ascribed to the points $x^{\alpha }$ of the Minkovski space, do not change
under deformation (for example see \cite{Shimbo95}):

- before deformation we can write 
\begin{equation}
\text{d}s^{2}=\eta _{\alpha \beta }\text{d}x^{\alpha }\text{d}x^{\beta } 
\end{equation}

- while after deformation the first order disturbances into the metric
tensor are: 
\begin{equation}
\text{d}S^{2}=g_{\alpha \beta }\text{d}X^{\alpha }\text{d}X^{\beta }\text{ ,
\ \ }h_{\alpha \beta }\thickapprox g_{\alpha \beta }-\eta _{\alpha \beta }%
\label{2.29}
\end{equation}%
For weak fields this formalism can lead us to equations for electromagnetic
and gravitation fields, when assuming that the disturbances $h_{\alpha \beta
}$ can be related to the following fields: 
\begin{itemize}
\item{$h_{\alpha \beta }^{G}$ --- the classical disturbances related to gravity}
\item{$h_{\alpha \beta }^{N}=N_{\alpha \beta }$ --- the disturbances related to EM field.}
\end{itemize}
We propose the metric tensor of the following form 
\begin{equation}
g_{\alpha \beta }=\eta _{\alpha \beta }+h_{\alpha
\beta }^{N}\ \ +h_{\alpha \beta }^{G} .\label{2.30}
\end{equation}

Introducing such first order disturbances, we can consider the
complex Riemann $ {R}_{\alpha \beta } $ and Einstein tensors
\begin{equation}
G_{\alpha \beta }= R_{\alpha \beta }-\frac{1}{2}g_{\alpha \beta } \label{2.31R}
\end{equation}
and the related basic field equation: 
\begin{equation}
{G}_{\alpha \beta }=0 \label{2.31G}
\end{equation}
Now we consider \eqref{2.31G} up to the first order terms.

\paragraph{Contributions from the non-diagonal terms $h_{\alpha \protect\beta }^{N}$.}

For non-diagonal terms we perturb the metric by the
potentials $\ h_{\alpha \beta }^{N}=N_{\alpha \beta}$  fulfiling the relation \eqref{2.8prime}.

 Considering the first order contributions to the Einstein tensor
we obtain (an index \emph {after} $|$ denotes differentiation)
\begin{equation}
G_{\alpha \beta }^{N}\thickapprox -( \frac{1}{2}N_{\mu \beta }\left\vert
_{\alpha }^{\mu }\right. +\frac{1}{2}N_{\mu \alpha }\left\vert _{\beta
}^{\mu }\right. ) + \frac{1}{2}N_{\alpha \beta }\left\vert _{\nu }^{\nu
}\right. =-\frac{1}{2}\text{\ }Y_{\alpha \beta }\ \label{2.33}
\end{equation}%
where accroding to \eqref{2.13} we have
$\frac{1}{2}N_{\mu \beta }\left\vert _{\alpha }^{\mu
}\right. +\frac{1}{2}N_{\mu \alpha }\left\vert _{\beta }^{\mu }\right. =%
\frac{1}{2c}( \eta _{\alpha ,\beta } + \eta _{\beta ,\alpha }).$ 

\paragraph{ Contributions from the h$_{\protect\alpha \protect\beta }^{G}$
terms}
The classical\ General Relativity relations 
\begin{equation}
G_{\alpha \beta }^{G}=R_{\alpha \beta }-1/2g_{\alpha \beta }=-\frac{8\pi G}{%
c^{4}}T_{\alpha \beta }\label{2.34} 
\end{equation}%
with the matter-energy tensor 
\begin{equation}
T_{\alpha \beta }=c^{2}\rho _{0}\upsilon _{\alpha }\upsilon _{\beta },\quad
\upsilon _{\alpha }=\{\upsilon _{s}/c, \ i \}  \label{2.34prime} 
\end{equation}
can be included into a new complex form, with the assumption that the
elements $h_{k4}^{G}$ present the imaginary values: 
\begin{equation}
\frac{1}{2}h_{\beta \alpha }^{G}\left\vert _{\nu }^{\nu }\right. +\frac{1}{2}%
\eta ^{\nu \mu }h_{\mu \nu }^{G}\left\vert _{\alpha \beta }\right.
\thickapprox \ \frac{8\pi G}{c^{4}}T_{\alpha \beta }\ \label{3.16}
\end{equation}

\paragraph{First order contributions to $G_{\protect\alpha \protect\beta }$.}

For the disturbances  given by equation \eqref{2.30} we obtain
\begin{equation}
G_{\alpha \beta }\thickapprox -\frac{1}{2}N_{\alpha \beta }\left| _{\nu
}^{\nu }\right. -\frac{1}{2}h_{\beta \alpha }^{G}\left| _{\nu }^{\nu
}\right. -\frac{1}{2}\eta ^{\nu \mu }h_{\mu \nu }^{G}\left| _{\alpha \beta
}\right. \label{4.9} 
\end{equation}
or if we define 
\begin{equation}
\bar{h}^{\mu \nu }=h^{\mu \nu }-\frac{1}{2}\eta ^{\mu \nu }h_{\alpha
}^{\alpha } 
\end{equation}
we arrive at
\begin{equation}
G_{\alpha \beta }\thickapprox -\frac{1}{2}N_{\alpha \beta }\left| _{\nu
}^{\nu }\right. -\frac{1}{2}\bar{h}_{\beta \alpha }^{G}\left| _{\nu }^{\nu
}\right. \thickapprox \frac{1}{c}\eta _{(\alpha ,\beta )}-\frac{1}{2}\
Y_{\alpha \beta }-\frac{8\pi G}{c^{4}}T_{\alpha \beta }.
\end{equation}
The matrics $Y_{\alpha \beta }$ is built also with the help of the $\epsilon$ matrices
\begin{equation}
Y_{\alpha \beta }=\tilde{Y}_{1}\epsilon ^{1}\ +\tilde{Y}_{2}\epsilon ^{2}+%
\text{i}\hat{Y}_{1}\epsilon ^{1}\ +\text{i}\hat{Y}_{2}\epsilon ^{2}
\end{equation}
while the tensor $\frac{1}{2c}(\eta _{\alpha ,\beta }+\eta _{\beta ,\alpha })
$ with the condition $\eta _{1}=\theta _{,1}$and $\eta _{2}=-\theta _{,2}$ ($%
\eta _{1,2}+\eta _{2,1}=0$ and $\eta _{1,2}-$ $\eta _{2,1}=2\theta _{,12}$ )
becomes  
\begin{equation*}
\frac{1}{2c}(\eta _{\alpha ,\beta }+\eta _{\beta ,\alpha })=\left[%
\begin{array}{cccc}
0 & 0 & \frac{1}{2}\eta _{1,3} & \frac{1}{2}\eta _{1,4} \\ 
0 & 0 & \frac{1}{2}\eta _{2,3} & \frac{1}{2}\eta _{2,4} \\ 
\frac{1}{2}\eta _{1,3} & \frac{1}{2}\eta _{2,3} & 0 & 0 \\ 
\frac{1}{2}\eta _{1,4} & \frac{1}{2}\eta _{2,4} & 0 & 0%
\end{array}%
\right]
\end{equation*}%
Finaly, we obtain 
\begin{equation*}
G_{\alpha \beta }\thickapprox \frac{1}{c}\eta _{(\alpha ,\beta )}-\frac{1}{2}%
\ Y_{\alpha \beta }-\frac{8\pi G}{c^{4}}T_{\alpha \beta }=\left[ 
\begin{array}{cccc}
0 & 0 & \text{i}\frac{1}{2}\hat{Y}^{2} & \frac{1}{2}\tilde{Y}_{1} \\ 
0 & 0 & -\text{i}\frac{1}{2}\hat{Y} & \frac{1}{2}\tilde{Y}_{2} \\ 
\text{i}\frac{1}{2}\hat{Y}^{2} & -\text{i}\frac{1}{2}\hat{Y}^{1} & 0 & 0 \\ 
\frac{1}{2}\tilde{Y}_{1} & \frac{1}{2}\tilde{Y}_{2} & 0 & 0%
\end{array}%
\right] -\frac{8\pi G}{c^{4}}T_{\alpha \beta }
\end{equation*}

Notice:\bigskip It is worth noticing a complimentary structures of the
amplitude related tensors $T_{\alpha \beta }$ and $E_{\alpha \beta }=(\frac{1%
}{2c}(\eta _{\alpha ,\beta }+\eta _{\beta ,\alpha })-\frac{1}{2}\ Y_{\alpha
\beta }):$ 
\begin{equation}
T_{\alpha \beta }=\left[ 
\begin{array}{cccc}
\text{Re} & \text{Re} & \text{Re} & \text{Im} \\ 
\text{Re} & \text{Re} & \text{Re} & \text{Im} \\ 
\text{Re} & \text{Re} & \text{Re} & \text{Im} \\ 
\text{Im} & \text{Im} & \text{Im} & \text{Re}%
\end{array}%
\right] \text{ },\text{ }E_{\alpha \beta }=\left[ 
\begin{array}{cccc}
0 & 0 & \text{Im} & \text{Re} \\ 
0 & 0 & \text{Im} & \text{Re} \\ 
\text{Im} & \text{Im} & 0 & 0 \\ 
\text{Re} & \text{Re} & 0 & 0%
\end{array}%
\right] 
\end{equation}
which implies a possibility of separation of the EM and gravity fields on a linear level. 

\paragraph{Remark: Meaning of natural potentials.}

Let us confine ourselves to the EM fields and let us consider the 3D
curvilinear complex space (or 6D real space) with coordinates 
\begin{equation}
\bar{X}_{s}=\tilde{X}_{s}+\text{i}\hat{X}_{s} . \label{3.17} 
\end{equation}
The potentials $\bar{A}_{s}=\tilde{A}_{s}+$i$\hat{A}_{s}$ can be identified
with such frames of the complex space $\bar{X}_{s}=\tilde{X}_{s}+$i$\hat{%
X}_{s}$;  same holds for the respecvtive tensors: $A_{\alpha \beta }\equiv 
\bar{X}_{\alpha \beta }:$%
\begin{equation}
\bar{X}_{s}=\bar{A}_{s}=\tilde{A}_{s}+i\hat{A}_{s}\text{ \ ; \ \ }\bar{X}%
_{\alpha \beta }\equiv \bar{A}_{\alpha \beta }\label{3.18} 
\end{equation}
We assume that at each point of this 6D complex space continuum  there can
appear the independent spin and twist motions\footnote{It can be
realized as "internal" motions of points in such "grained" (micropolar) 
space based on the Planck length.}
and that complex space is combined to the point of such continuum
\begin{equation}
\epsilon _{abs}\bar{X}_{s}=\epsilon _{abs}\tilde{X}_{s}+\text{i}\epsilon
_{abs}\hat{X}_{s}\text{ , \ \ \ }\bar{A}_{s,s}=\text{\ }\bar{X}_{s,s}=%
\tilde{X}_{s,s}+\text{i}\hat{X}_{s,s}=0 . \label{3.19} 
\end{equation}

\section*{Conclusion and perspectives }

In this initial work we have presented the Complex Relativity theory as inspired by    
recent progress in asymmeric continuum approach to material sciences. 
The goal of this work is to propose the common relativity framework 
for electromagnetic and gravity fields as close as possible 
to the classical General Relativity formulation.
For another classical approach see \cite{Sachs}.

We mention here possible ways of further development of our approach. 

The relations for the symmetric tensors as expressed by the $\gamma$ -tensors
remain valid in any reference system, but we might also return to the system with the antisymmetric
potentials ($\bar{A}^{3}$ different from zero; see relations between natural
potentials in both systems) in order to get notation comparable with the
gravity part. We  shall be also aware that in such a case we return to
antisymmetric natural potential tensor; we would obtain\ 
\begin{equation*}
G_{\alpha \beta }^{ANTISYM} = \frac{1}{2c} \times 
\end{equation*}
\begin{equation}
\times \left[ 
\begin{array}{llll}
0 & \epsilon _{3dk}\psi _{k,d}-\text{i}\frac{\partial }{\partial t}\psi _{3}
& -\epsilon _{2dk}\psi _{k,d}+\text{i}\frac{\partial }{\partial t}\psi _{2}
& -\epsilon _{1dk}\psi _{k,d}+\text{i}\frac{\partial }{\partial t}\psi _{1}
\\ 
-\epsilon _{3dk}\psi _{k,d}+\text{i}\frac{\partial }{\partial t}\psi _{3} & 0
& \epsilon _{1dk}\psi _{k,d}-\text{i}\frac{\partial }{\partial t}\psi _{1} & 
-\epsilon _{2dk}\psi _{k,d}+\text{i}\frac{\partial }{\partial t}\psi _{2} \\ 
\epsilon _{2dk}\psi _{k,d}-\text{i}\frac{\partial }{\partial t}\psi _{2} & 
-\epsilon _{1dk}\psi _{k,d}+\text{i}\frac{\partial }{\partial t}\psi _{1} & 0
& -\epsilon _{3dk}\psi _{k,d}+\text{i}\frac{\partial }{\partial t}\psi _{3}
\\ 
\epsilon _{1dk}\psi _{k,d}-\text{i}\frac{\partial }{\partial t}\psi _{1} & 
\epsilon _{2dk}\psi _{k,d}-\text{i}\frac{\partial }{\partial t}\psi _{2} & 
\epsilon _{3dk}\psi _{k,d}-\text{i}\frac{\partial }{\partial t}\psi _{3} & 0%
\end{array}%
\right] \ 
\end{equation}%
where $G_{\alpha \beta }^{ANTISYM}$ is no longer the complex Einstein
tensor, but stands for the corresponding antisymmmetric expression.

In the above considerations we tried to preserve the symmetric property when
constructing the Complex Relativity, however, it seems more natural to
admit the possibility that the perturbations into metric and to Einstein
tensor can be asymmetric or even antisymmetric. 

These problems will be discussed in our next paper.

\section*{Acknowledgments}
The authors thank Andrzej Czechowski for useful remarks.
This work is partially supported by Polish Ministry of Science and Information Society Technologies,  
project 2PO4D 060 28.

\bibliographystyle{amsplain}
\bibliography{mybib}



\end{document}